\documentclass[%
aip,
 amsmath,amssymb,
 reprint,%
]{revtex4-1}

\usepackage{graphicx}% Include figure files
\usepackage{dcolumn}% Align table columns on decimal point
\usepackage{bm}% bold math
\usepackage[utf8]{inputenc}
\usepackage[T1]{fontenc}
\usepackage{mathptmx}
\usepackage{etoolbox}
\usepackage{color}
\usepackage[percent]{overpic}
\definecolor{green2}{RGB}{0,115,4}
\usepackage{hyperref}

\makeatletter
\def\@email#1#2{%
 \endgroup
 \patchcmd{\titleblock@produce}
  {\frontmatter@RRAPformat}
  {\frontmatter@RRAPformat{\produce@RRAP{*#1\href{mailto:#2}{#2}}}\frontmatter@RRAPformat}
  {}{}
}%
\makeatother
\begin{document}

\title[]{The effects of hydrodynamic interactions 
\textcolor{black}{on} 
the swimming velocity and stability of a swarm of microswimmers}
% Force line breaks with \\
\author{Andr\'{e}s~C\'{o}rdoba}
\affiliation{Pritzker School of Molecular Engineering, University of Chicago, 
5640 South Ellis Avenue Chicago, IL 60637, USA}
\author{Jay~D.~Schieber}%
\affiliation{Center for Molecular Study of Condensed Soft Matter, Department of Chemical and Biological Engineering, Department of Physics, and Department of Applied Mathematics, Illinois Institute of Technology, 3440 South Dearborn Street, Chicago, IL 60616, USA}

\date{\today}% It is always \today, today,

\begin{abstract}
The dynamics and stability of a swarm of microswimmers is examined here 
using a thermodynamically compliant microswimmer model. 
The theory presented here is a mean-field model in which
the swarm is considered to be a uniform solution of swimmers 
that are moving, on average, in the same direction. 
The hydrodynamic interaction between swimmers is modeled accurately 
assuming a crystalline arrangement to the swarm. A swimmer in the swarm can 
swim up to twelve times faster than when alone in an infinite sea,
when the volume fraction of swimmers in the swarm is about 0.14.
Moreover we have also studied the stability of the swarm 
\textcolor{black}{around a uniform crystalline arrangement}
by calculating the hydrodynamic torques generated by the 
swarm as a function of the volume fraction of swimmers. 
The predictions presented here agree with recent 
multiparticle simulations that have shown that hydrodynamic 
torques have a stabilizing effect in swarms of pullers while swarms of 
pushers are generally destabilized by hydrodynamic 
interactions. The thermodynamically admissible coupling  between
the swimmer's motion and fuel consumption allows us to study the full dynamics
instead of artificially constrained steady-states only.
By accounting for fuel consumption and high order 
hydrodynamic interactions we are able to examine
the swarm's stability as function of fuel concentration and the 
volume fraction of swimmers. We find that at high 
concentrations of fuel, swarms of pullers are stabilized by hydrodynamic
torques for volume fractions of swimmers as low as 0.02 but at lower volume fractions 
Brownian forces make the swarm unstable.

\par\medskip
\noindent\textit{This article may be downloaded for personal use only. Any other use requires prior permission of the author and AIP Publishing. This article appeared in Phys. Fluids 35, 111901 (2023)
and may be found at \url{https://doi.org/10.1063/5.0174558}}

\end{abstract}

\maketitle

\section{\label{sec:level1}Introduction}

Microswimmers can be described broadly as 
micrometer-sized particles that can convert local chemical energy into propulsive forces.
This allows microswimmers to move at a constant velocity 
through their surrounding fluid medium.
Microswimmers are common in nature, for example 
swimming microorganisms, which include bacteria, algae, and spermatozoa, 
play a fundamental role in most biological processes.  While the size, shape, and
propulsion mechanism are diverse, all microswimmers
swim at small Reynolds numbers  \cite{OyamaJPSJ2017, gomez2017tuning, 
LugliJPCC2012}. Microorganisms are able to overcome the Brownian forces of 
their surroundings by harvesting energy to navigate in viscous fluid environments.
Moreover in moderately concentrated
suspensions microswimmers exhibit complex collective behavior 
including swarming and self-assembly \cite{krishnamurthy2015collective,
wu2000particle,sokolov2007concentration, dunkel2013fluid,wu2006collective}.

\textcolor{black}{
For example a wide variety of bacteria exploit
helical filaments, called flagella, for their propulsion. 
Bacteria like \textit{Escherichia coli} and \textit{Salmonella} swim in a 
``run-and-tumble'' motion \cite{elgeti2015physics, schwarz2016escherichia}. 
In the ``run'' phase the flagella form a bundle and the bacterium moves forward
in a direction determined by its long axis. 
At the beginning of the ``tumble'' phase, one flagellum reverses its rotational 
direction. This flagellum leaves
the bundle which implies a random reorientation of the bacterium 
\cite{elgeti2015physics, schwarz2016escherichia}.
Bacteria not only swim as individuals, but exhibit collective behavior at a 
moist surface or in a thin liquid film in the form of swarming \cite{partridge2013swarming}.
Eucaryotic swimmers are usually
propelled by cilia. These motile hair-like extensions of the cell consist of a 
bundle of microtubules. 
\textit{Chlamydomonas reinhardtii} is a single-celled green algae of about 10 $\mu$m.
The swimming motion of \textit{Chlamydomonas}
resembles the human breast-stroke: the flagella are pulled back
in a nearly straight shape, and are then bend over and pushed
forward again \cite{elgeti2015physics, kantsler2013ciliary}.}

Synthetic colloidal microswimmers are capable of mimicking complex 
biolocomotion by means of simple self-propulsion mechanisms.
A widely studied type of artificial microswimmers are Janus 
particles, which as the name indicates have two distinct parts, an inert half and 
a catalytic half. The catalytic part degrades fuel molecules, leading to localized concentration
gradients of the reactants around the Janus particle \cite{anderson1989colloid} .
Janus particles have been observed to swim in the direction of this 
self-generated concentration gradients \cite{EbbensLangmuir2011,sharan2022upstream}.
Artificial self-propelled micro and nanoengines, or swimmers, have attracted the
interest of experimental and theoretical researchers. 
Synthetic microswimmers also work at low
Reynolds number, where inertia does not sustain motion 
once the driving force stops, and often
require complicated chemistry in both structural and dynamical terms \cite{LugliJPCC2012}. 
Synthetic microswimmers have been extensively 
studied experimentally using microscopy techniques 
\cite{EbbensLangmuir2011, gomez2017tuning, Michelin2015, Wilson2013}. 
Janus particle swimmers made by
coating fluorescent polymer beads with hemispheres of platinum 
have been fully characterized using video microscopy to
reveal that they undergo propulsion in hydrogen peroxide fuel
away from the catalytic platinum patch \cite{EbbensLangmuir2011}.
Among the few methods which have been proposed to create small-scale swimmers, 
those relying on self-phoretic mechanisms present an interesting design 
challenge in that chemical gradients are required to generate net propulsion \cite{Michelin2015}. 
Asymmetries in geometry are sufficient
to induce chemical gradients and swimming  \cite{Michelin2015}. 
Geometric asymmetries can be tuned to induce large
locomotion speeds without the need of chemical patterning \cite{Michelin2015}.

Microswimmers have also been studied theoretically and through computer simulations 
\cite{Datt2018, Yasuda2017, Nasouri2017, Milster2017, 
childressJFM2012,SabassJCP2012, FalascoPRE2016, 
golestanian2005propulsion,
popescu2010phoretic,thakur2011dynamics}.
For instance, it has been shown that in catalytic Janus particles
the catalytic reaction produces an asymmetric, 
non-equilibrium distribution of reaction products
around the colloid, which generates osmotic or other
phoretic forces \cite{golestanian2005propulsion, popescu2010phoretic, 
SabassJCP2012,thakur2011dynamics, LugliJPCC2012}. 
Analogous to the behavior of macroscopic motors the particle 
shape matters and influences not only the
random diffusion but also the efficiency of the propulsive motion \cite{LugliJPCC2012}.
Coarser models of chemically propelled Janus swimmers have also
been previously proposed \cite{Milster2017,OyamaJPSJ2017}.
For example, the dynamical description of the 
motion of a stochastic microswimmer with constant speed and fluctuating
orientation in the long time limit has been 
obtained by adiabatic elimination of the orientational variable \cite{Milster2017}. 
Other simulations have used the squirmer
model \cite{guan2023squirmers}, 
which provides an idealized representation of swimmers as spheroidal particles 
that propel owing to a modified boundary condition at their surface.
With this type of model the single-particle and many-particle dynamics of swimmers in
bulk and confined systems have been studied using the smoothed profile method, 
which allows efficient solving of the coupled particle-fluid problem  \cite{OyamaJPSJ2017}.

Another line of work focuses on the collective behavior that 
emerges in moderately concentrated
suspensions of microswimmers \cite{krishnamurthy2015collective,
saintillan2008instabilities, hernandez2009dynamics, simha2002hydrodynamic,
pawan2022active}.
In those works individual swimmers  are usually modeled using force
dipoles. The hydrodynamic interactions
between swimmers and/or confining walls are treated 
using Stokesian dynamics \cite{brady1988stokesian} 
or similar methods \cite{hernandez2007fast}.
Other simulation technique that has been 
used to study the collective dynamics of Janus microswimmers 
combines molecular dynamics with multiparticle 
collision dynamics. Each Janus motor is modeled as a hard sphere and solvent 
molecules undergo reactive and nonreactive collisions with it \cite{colberg2017many,
huang2017chemotactic}.
Experimentally, this ``collective behavior'' has been observed
in moderately concentrated suspensions of swimming \textit{Escherichia coli}
\cite{wu2000particle} and \textit{Bacillus subtilis} \cite{sokolov2007concentration, dunkel2013fluid}.
Other related observations include swimming speeds much greater than that of an isolated
swimmer \cite{sokolov2007concentration}, enhanced swimmer and tracer diffusivities
\cite{wu2000particle, wu2006collective} and efficient fluid mixing
\cite{sokolov2009enhanced}. A common feature of the aforementioned
experiments is that they all involve rear-actuated swimmers known as \textit{pushers}.

Other theoretical work has focused on the details of the interaction between 
pairs of swimmers swimming very close to each other. 
For instance the squirmer model and a particle-based simulation approach
was employed to study the swimming behavior of 
microswimmer dumbbells \cite{clopes2020hydrodynamic}.
Using a perturbative analytic approach, scaling laws and
estimates of the strength of the interactions in the case of a pair of 
three-sphere swimmers have also 
been recently proposed \cite{ziegler2021theoretical}. 
Using that analytic approach it has been shown that for pairs 
of swimmers, the swimming velocity increases when compared to
individual swimmers, even when the swimmers are one behind another 
\cite{ziegler2021theoretical}. More recently the interaction and motion 
of a pair of squirmers swimming in shear flow was studied using the lattice Boltzmann method 
\cite{guan2023squirmers}.

As mentioned above, one of the main question that arises when multiple microswimmers interact 
with each other is the stability of these microswimmer suspensions or swarms.
The microswimmers in a swarm can change their orientations due to hydrodynamic 
and steric (short-range repulsive) interactions
with other swimmers. Some microswimmers also
have intrinsic orientation relaxation mechanisms. \textcolor{black}{As discussed above, 
an example of these are bacteria such as \textit{Escherichia coli} in which the
reorientation mechanism is related to the way their
flagella function and execute a run-and-tumble motion. However
the normal run-tumble behavior seen in swimming of isolated bacteria
is significantly suppressed in swarms \cite{partridge2013swarming}. The cells attempt to
move straight ahead but are continuously jostled either from
true collisions with neighbors or from forces induced by the 
moving fluid \cite{darnton2010dynamics, partridge2013swarming}.}
Swimmers that lack an intrinsic orientation relaxation
mechanisms, and that change their orientation only due to hydrodynamic interactions,
are usually referred to as straight swimmers in the literature. 

The stability of microswimmer suspensions has been studied previously using both
mean-field models and multiparticle numerical 
simulations \cite{simha2002hydrodynamic,saintillan2008instabilities, 
saintillan2008instabilitiesPF,hohenegger2010stability,
underhill2008diffusion,evans2011orientational}. 
However, previous mean-field
kinetic theory models of microswimmer swarms have neglected
the contributions from particle hydrodynamic interactions and only the lowest order 
contribution from the single-particle swimming has been included \cite{saintillan2008instabilitiesPF}.
An exception is the work of Subramanian and Koch \cite{subramanian2009critical} which did account
for hydrodynamic interactions and intrinsic orientation relaxation mechanisms. 
However that work used an averaged treatment 
of the hydrodynamic interactions and did not consider the stability around a uniform
\textcolor{black}{crystalline arrangement} as we do in this paper. More recent work 
introduced particle-based simulations of suspensions of hydrodynamically
interacting swimmers with intrinsic orientation relaxation mechanisms 
\cite{krishnamurthy2015collective}.
However, previous kinetic theories and simulations of microswimmer swarms
have not included
the coupling between the swimming dynamics and fuel consumption and therefore 
do not allow for a thermodynamically compliant description of these systems.

Microswimmers operate far from equilibrium, therefore a complete
and rigorous thermodynamic analysis of systems with microswimmers requires
a thermodynamics framework capable of handling systems far from equilibrium.
GENERIC (general equation for the nonequilibrium 
reversible-irreversible coupling) is an equation to describe the time-evolution
of the state variables not only near equilibrium but far from the
equilibrium state \cite{ottinger2005beyond, 
grmela1997dynamics, ottinger1997dynamics}. 
GENERIC comprises reversible and irreversible 
contributions to the entire dynamics of the system. The
reversible or mechanical contribution is driven by the gradient
of the system energy with respect to the state variables, whereas
the irreversible or dissipative dynamics is driven by the thermodynamic 
forces represented by the gradient of the entropy.
Thus the two generators, the total energy and the total entropy
of the system, drive the full dynamics of the system 
represented by the state variables.

Recently we introduced a mean-field model for a single microswimmer
\cite{cordoba2020simple}. The model was inspired by the
nonequilibrium thermodynamics of multi-component fluids that undergo chemical
reactions. These thermodynamics can be rigorously described
in the context of the GENERIC framework \cite{IndeiGENERIC2017}.
The swimmer model considered here 
consists of a polymer dumbbell in a sea of reactants. One of the beads of the dumbbell 
is allowed to act as a catalyst for a chemical reaction between the reactants.
The mean-field Janus
swimmer model has a small number of adjustable parameters. 
These are related to the diffusion coefficients and thermodynamic
derivatives of the chemical potentials of the reactants and the dumbbell, the
rate constant for the surface reaction in the catalytic bead, 
the size of the beads and the rest length of the dumbbell.
Moreover, the entropy balance for the model allows
one to check that the entropy production rate stays positive for a given
set of parameters. In this way it can be guaranteed that 
the the second law of thermodynamics is always 
obeyed in simulations using the model. The 
thermodynamically admissible coupling between fuel consumption 
and swimmer motion allows us to study the full dynamics
instead of artificially constrained steady-states only.

In this paper the thermodynamically compliant mean-field microswimmer model
is extended to study swarms of microswimmers.
By a swarm here we mean a uniform solution of swimmers 
that are moving, on average, in the same direction. 
We do not attempt here to describe the emergence of such ordered sate, we only 
consider its stability. We also consider 
the swarm to be very large and do not attempt to account for the effect 
of boundaries at the edge of it. In general our theoretical framework should be 
an accurate description of the dynamics of swimmers
in a swarm far from the edges of it. In particular we calculate the velocity of the swimmers 
in the swarm as a function of the volume fraction of swimmers.
Moreover the stability of the swarm
is analyzed as a function of the volume fraction of swimmers in the swarm.
Differently from most of the previous mean-field models of microswimmer
swarms we account for high order hydrodynamic interactions between microswimmers.
This allows us to study the swarm's stability as function of the 
volume fraction of swimmers.
The remainder of this paper is organized as follows: 
In Section \ref{model} we describe how our
mean-field microswimmer model \cite{cordoba2020simple} can be 
used to describe swarms of microswimmers. In particular,
we derive the translational and rotational swimming velocities of
a microswimmer in a \textcolor{black}{crystalline} periodic array of other microswimmers.
In the Section \ref{results}, we include thermal fluctuations and 
perform a stability analysis of the swarm as a function of the volume fraction
of microswimmers. Finally some
concluding remarks are given in Section \ref{conclusions}.

\section{Model and methods} \label{model}

\begin{figure}[h t]
\center
\includegraphics[width=\linewidth]{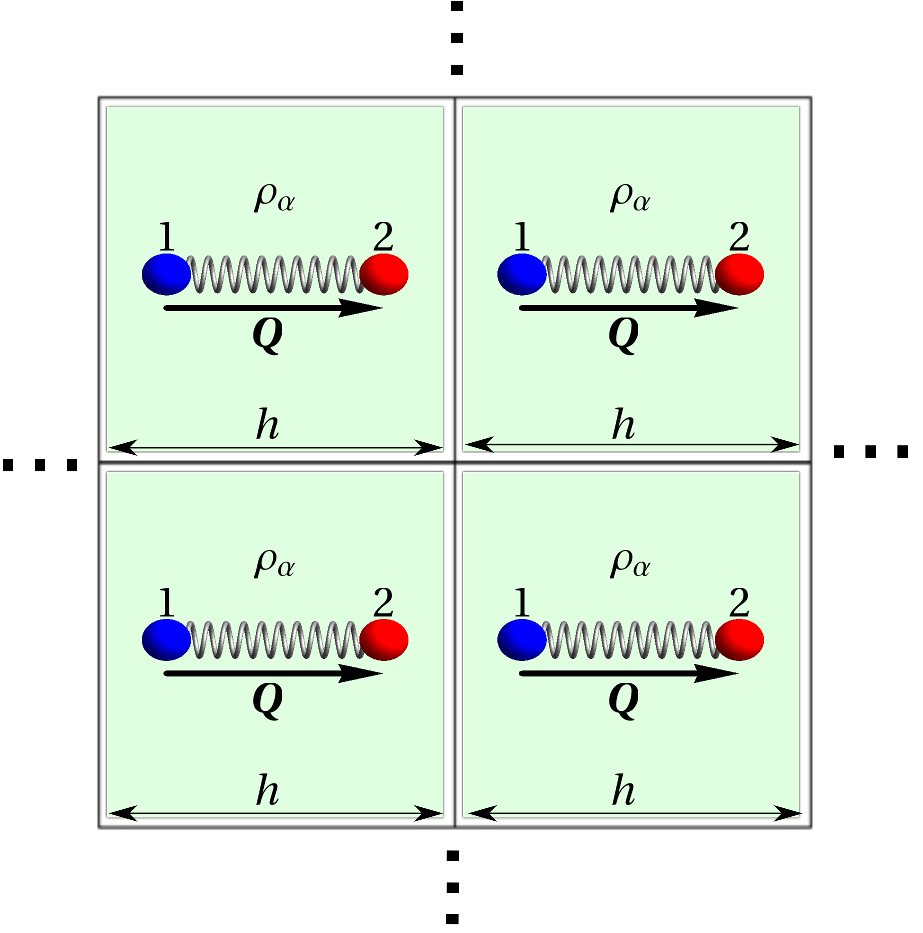}
\caption{
Sketch of a section 
of the periodic array of  the Janus dumbbell swimmers considered in this work.
The position of bead $1$ is given by $\bm{r}_1$ and the position 
of bead $2$ is given by $\bm{r}_2$, the following definitions
are also used throughout the text,
$\bm{Q}:=\bm{r}_{2}-\bm{r}_{1}$ and $\bm{r}_c:=\dfrac{\bm{r}_1+\bm{r}_2}{2}$.
The radius of both beads is $R$. Bead $1$
acts as a catalyst for the irreversible chemical reaction $|\nu_\alpha|\alpha  
\longrightarrow |\nu_\beta|\beta$.
The swimmer is immersed in a bath of reactants and
the average concentration of reactant, is equal to $\rho_{\alpha}$.
The unit cell of the array has length $h$ and volume $V_0=h^3$. }
\label{fig3}
\end{figure}

The swimmer model considered here is a bead-spring dumbbell with 
the position of bead $1$ given by $\bm{r}_1$ and the position 
of bead $2$ given by $\bm{r}_2$. The bond vector connecting the two beads is
$\bm{Q}:=\bm{r}_{2}-\bm{r}_{1}$, and the 
center of mass of the dumbbell is $\bm{r}_c:=\dfrac{\bm{r}_1+\bm{r}_2}{2}$.
A sketch of the dumbbells is shown in Fig. \ref{fig3}.
The inertia of the beads is neglected and over-damped dynamics are considered
for the equations of motion of the dumbbell. The dumbbell is dissolved in a 
solvent formed by species $\alpha$ and $\beta$. Bead $1$ of the dumbbell
acts as a catalyst for the irreversible chemical reaction $|\nu_\alpha|\alpha  
\longrightarrow |\nu_\beta|\beta$, where 
$\nu_\alpha$ and $\nu_\beta$ are stoichiometric coefficients.

The swimmer is immersed in a bath of reactants and
the average concentration of reactant in the bulk 
is equal to $\rho_{\alpha}$. Note that $\rho_{\alpha}$ will be allowed to 
change with time, but it will not be a field variable.
The level of description (set of state variables) of the model is,
$\bm{x}(\bm{r}_c,\bm{Q})=\left\{ \rho_{\alpha}, \rho_c(\bm{r}_c,\bm{Q}), 
\epsilon(\bm{r}_c)\right\}$. Where
$\rho_c(\bm{r}_c,\bm{Q})$ is the mass density 
of dumbbells and $\epsilon(\bm{r}_c)$ is the internal energy density.
Note that $\dfrac{\rho_c(\bm{r}_c,\bm{Q})}{\overline{\rho}_c}$ can be interpreted as the probability 
density of finding a dumbbell with its center of mass at position $\bm{r}_c$ and
end-to-end vector $\bm{Q}$.
With $\overline{\rho}_c:= 
\dfrac{\int \int \rho_c(\bm{r}_c,\bm{Q})d\bm{Q}d\bm{r}_c}{V_0}=\dfrac{m_c}{V_0}$
being the normalization factor for the probability density 
and $V_0$ is the volume of the bath.
Moreover it is assumed that very near bead $1$ of the dumbbell, 
{\it i.e.} the catalytic bead, the concentration of species $\alpha$ 
is always very small and is approximated to be always zero. 
At a distance $\ell_0$ away, near bead $2$ of the dumbbell,
the concentration of species $\alpha$ is assumed
to always be equal to $\rho_{\alpha}$,
therefore the concentration gradient of $\alpha$ is
$\dfrac{\partial \rho_\alpha(\bm{r}_c)}{\partial \bm{r}_c}\approx
\dfrac{\bm{Q} }{Q} \dfrac{\rho_{\alpha}}{\ell_0} 
\dfrac{\rho_c(\bm{r}_c,\bm{Q})}{\overline{\rho}_c}$.
It has been previously shown \cite{cordoba2020simple}
that the  stochastic differential equations that describe the 
motion of this Janus swimmer are,
\begin{align}\label{modelf}
d\bm{Q}=&\frac{2\gamma'_{cc}\left(D'_{c\alpha}+D'_{c\beta}\right)}{k_{\rm B} T}
\frac{\partial \phi^{(S)}}{\partial \bm{Q}}dt \\ \nonumber
+&\sqrt{4\gamma'_{cc}\left(D'_{c\alpha}+D'_{c\beta}\right)}
d\bm{W}_1(t),\\ \nonumber
d\bm{r}_c=&\left[\frac{\rho_{\alpha}(t)}{\ell_0 \overline{\rho}_c}
\left(D'_{c\alpha}\gamma'_\alpha+
D'_{c\beta}\gamma'_\beta\right)\frac{\bm{Q}}{Q}\right]dt \\ \nonumber
+&\sqrt{ D'_{c\alpha} \gamma'_{c\alpha}+
D'_{c\beta} \gamma'_{c\beta}}d\bm{W}_2(t).
\end{align}
Where $d\bm{W}_1$ and $d\bm{W}_2$ are vectors of Wiener increments
with statistics, $\left\langle {d \bm{W}}_{i}(t)\right\rangle_{\rm eq}=\bm{0}$ and 
$\left\langle {d\bm{W}}_{i}(t){d\bm{W}}_{j}(t')\right\rangle_{\rm eq}
=\bm{\delta}\delta(t-t')\delta_{ij}dtdt'$. Where $\delta(...)$ is the Dirac delta 
function, $\bm{\delta}$ is the identity tensor, and $\delta_{ij}$ is the Kronecker
delta. In what follows it will be assumed that the potential energy of the elastic 
dumbbell is a Fraenkel spring $\phi^{(S)}=-\dfrac{k_b (Q-\ell_0)^2}{2}$, 
with $k_b$ the entropic spring constant.
$D'_{c\alpha}$ and $D'_{c\beta}$ are diffusion coefficients 
and $\gamma'_{cc}$, $\gamma'_\alpha$, $\gamma'_\beta$,
$\gamma'_{c\alpha}$ and $\gamma'_{c\beta}$, are 
thermodynamic derivatives of the chemical potentials of the reactants and the
swimmer, \textcolor{black}{
\textit{e.g.} $\gamma_{cc}'=\dfrac{1}{T}\left[\dfrac{\partial \mu_c(\bm{r},\bm{Q})}
{\partial \rho_c(\bm{r},\bm{Q})}\right]_{T,p,\rho_\alpha}$. Where $\mu_c(\bm{r},\bm{Q})$ 
is the chemical potential of the swimmer, $T$ is the temperature and 
$p$ is the pressure. The definitions of $\gamma'_\alpha$ and $\gamma'_\beta$
involve derivatives of the chemical potentials of the reactants and the swimmer 
with respect to the concentration of reactants \cite{cordoba2020simple}. 
Moreover the definitions of 
$\gamma'_{c\alpha}$ and $\gamma'_{c\beta}$
involve derivatives of the chemical potentials of the reactants and the swimmer 
with respect to  $\rho_c(\bm{r},\bm{Q})$.
These thermodynamic derivatives are used as adjustable 
parameters, but as explained below the second law 
of thermodynamics imposes some restrictions on the values that they can take.}

For the level of description of the mean-field 
swimmer model one can write an evolution equation
for the concentration of reactant,
\begin{equation}\label{fuelc}
\frac{d\rho_{\alpha}(t)}{dt}=-J(t).
\end{equation}
Where the net reaction rate $J(t)$ includes both 
the mass transport of reactants from the bulk
to the dumbbell and also the 
reaction rate once the reactants reach the catalyst-coated 
bead of the Janus dumbbell.  
Therefore the net reaction rate, $J(t)$, can be written as,
\begin{align}\label{jrate}
J(t)=\left(\frac{4 \pi R^2}{V_0}\right)
\frac{k'' \gamma_{\alpha \alpha \beta} D'_{\alpha\alpha}
\rho_{\alpha}(t)}
{\gamma_{\alpha \alpha \beta} D'_{\alpha\alpha}+\ell_0 k''}.
\end{align}
$R$ is the radius of the 
catalyst-coated bead and again $V_0$ is the volume of the 
bath where the Janus dumbbell is swimming.
The rate constant for the reaction at the surface of the
catalyst-coated bead of the Janus dumbbell is $k''$.
The specific form for the reaction rate used in eq. (\ref{jrate})
is based on the solution to the problem of
binary diffusion through a boundary layer and 
reaction at a surface \cite{bird2007transport, cordoba2020simple}.

Note that in eq. (\ref{modelf}) the first term on the 
right hand side of the equation for $\bm{r}_c(t)$
can be interpreted as the swimming velocity
for a single swimmer in an infinite sea,
$V^{\rm S}=\dfrac{\rho_{\alpha}}{\ell_0 \overline{\rho}_c}
\left(D'_{c\alpha}\gamma'_\alpha+
D'_{c\beta}\gamma'_\beta\right)$. 
Note that the dumbbell swims in the direction 
$\bm{Q}/Q$ which as was mentioned above is the direction
of the concentration gradient of $\alpha$. $V^{\rm S}$ is
directly proportional to the concentration of reactant in the bath, $\rho_{\alpha}$, 
and  inversely proportional to the rest length of the dumbbell, $\ell_0$. 
This dependence of $V^{\rm S}$ on $\rho_{\alpha}$ and $\ell_0$ 
also arises from the form of $\dfrac{\partial \rho_\alpha(\bm{r}_c)}{\partial \bm{r}_c}$
given above. Note that shorter 
dumbbells produce larger concentrations gradients around them, which
in turn produces a larger swimming velocity.

In the GENERIC framework the entropy generation rate 
density of a model can be calculated as previously shown \cite{cordoba2020simple}.
In general, the entropy generation rate should always be positive
to guarantee that the model satisfies the second law of thermodynamics.
For example, in the simplified case when $\gamma'_\beta=\gamma'_\alpha$
and  $\gamma'_{c\beta}=\gamma'_{c\alpha}$ the restrictions on the model 
parameters found from the entropy balance simplify to,
$\gamma'_\alpha>0, ~ \gamma'_{c\alpha}>0$.
In what follows these restrictions are used
to guide the selection of parameters when performing numerical 
simulations of the model. Note that in the level 
of description of our model 
the details of the swimming mechanism
are contained in the parameters  $D'_{c\alpha}$, $\gamma'_\alpha$, 
$D'_{c\beta}$ and $\gamma'_\beta$. The values of these parameters for a specific
system can be obtained from more detailed models or simulations or by
fitting to experimental data. But as mentioned above
the second law of thermodynamics imposes some restrictions on the values 
that $\gamma'_\alpha$, $\gamma'_\beta$, $\gamma'_{c\beta}$ and $
\gamma'_{c\alpha}$ can take.

\textcolor{black}{In this work, using the same 
mean-field level of description described above we 
propose a model to describe a swarm of microswimmers.
In particular we seek to study the effect that hydrodynamic interactions
between swimmers have in the swimming velocity and stability of the swarm. 
At this level of description the problem then reduces to
finding the effect that other swimmers in the swarm may have in the 
forces that act on $\bm{r}_c$ and $\bm{Q}$. To accomplish this here 
we focus} on a swarm that is initially in a 
\textcolor{black}{uniform crystalline arrangement}. 
The stability around that particular state is examined.
The  stochastic differential equations that describe the 
motion of the Janus swimmer in the swarm can be written as,
\begin{align}\label{SDEs}
d\bm{Q}=& \left[\bm{V}_{\rm conf}^{\rm S}+
\frac{2\gamma'_{cc}\left(D'_{c\alpha}+D'_{c\beta}\right)}{k_{\rm B} T} 
\frac{\partial \phi^{(S)}}{\partial \bm{Q}}\right]dt \\ \nonumber
&+\sqrt{4\gamma'_{cc}\left(D'_{c\alpha}+D'_{c\beta}\right)}
d\bm{W}_1(t),\\ \nonumber
d\bm{r}_c=&\bm{V}_{\rm cm}^{\rm S}dt+
\sqrt{ D'_{c\alpha} \gamma'_{c\alpha}+
D'_{c\beta} \gamma'_{c\beta}}d\bm{W}_2(t).
\end{align}
Note that compared to eq. (\ref{modelf}), eq. (\ref{SDEs}) contains additional
or modified terms on the right hand side of the evolution equations for $\bm{Q}$ 
and $\bm{r}_c$, 
namely $\bm{V}_{\rm conf}^{\rm S}$ and $\bm{V}_{\rm cm}^{\rm S}$. 
These new terms include the effects of hydrodynamic interactions
with others swimmers in the swarm on the dumbbell's position and conformation vectors.
To find these contributions we start with
the Stokes equation and the mass balance for an incompressible solvent,
\begin{align}\label{stokes_real}
\mu \nabla^2 \bm{v}(\bm{r})&=\nabla p(\bm{r}) + \sum_c
\bm{F}\left(\bm{r},\bm{r}_c, \bm{Q}\right),  \\ \nonumber
\nabla \cdot \bm{v}(\bm{r})&=0.
\end{align}
where $\mu$ is the viscosity of the fluid where the swarm is immersed 
$\bm{v}(\bm{r})$ is the velocity field in the fluid and $p(\bm{r})$ is the pressure field in 
the fluid. $\bm{F}\left(\bm{r},\bm{r}_c, \bm{Q}\right)$
is the force density exerted on the fluid at position $\bm{r}$
by the center of mass of a
Janus dumbbell located at position  $\bm{r}_c$ and with
end-to-end vector $\bm{Q}$. The sum in eq. (\ref{stokes_real}) is over 
all the swimmers in the swarm. Assuming that the force density that 
the swimmer exerts on the fluid can be described by a function that
varies only along  $\bm{Q}$ (\textit{i.e.} an axisymmetric swimmer), 
then
\begin{align}
\bm{F}\left(\bm{r},\bm{r}_c, \bm{Q}\right)=
- \dfrac{\bm{Q}}{Q} 
\int_{-\infty}^{\infty} \delta\left(\bm{r}-\bm{r}_c-s\dfrac{\bm{Q}}{Q}\right)
f(s)ds.
\end{align}
Where $s$ is a coordinate 
that runs along the swimmer's length in the direction of $\bm{Q}$, 
$f(s)$ is the force density distribution
along the swimmer's length and $\delta(...)$
is the Dirac delta function. Consistent with our 
treatment of the swimmer as a dumbbell we split the calculation of 
$\bm{F}(\bm{r},\bm{r}_c,\bm{Q})$ into two parts. We first calculate the
effect of the hydrodynamic forces on the center of mass of the dumbbell,
\textcolor{black}{$\bm{F}_{\rm cm}(\bm{r},\bm{r}_c,\bm{Q})$}.
Since the swimmers are moving, for this first 
calculation the key part is to use an inertial frame of reference.
In such frame of reference the conservation of linear momentum 
is obeyed. 
Therefore the calculation properly accounts for the transfer of linear momentum 
from the swarm of swimmers to the fluid. 
In a second part we calculate the effect of the hydrodynamic torques on the 
orientation of the swimmer, \textcolor{black}{$\bm{F}_{\rm conf}(\bm{r},\bm{r}_c,\bm{Q})$}. 
For that second calculation some 
details of the swimming mechanism 
become relevant, these details are discussed below.

We will start by calculating the velocity of the center of mass of a 
Janus dumbbell in the swarm, $\bm{V}_{\rm cm}^{\rm S}$.
For this calculation a first-order approximation of $f(s)$
is sufficient. Note that the first order approximation does 
not allow distinguishing between different swimming mechanisms
(\textit{e.g.} pullers or pusher).
Therefore we will use $f(s)\approx\zeta_{\rm cm} V^{\rm S}\delta(s)$,
where $V^{\rm S}$ is the swimming velocity of a single swimmer 
in an infinite sea given above, and $\zeta_{\rm cm}$ is a friction 
coefficient. Therefore the force density becomes, 
\begin{align}\label{forcecmr}
\bm{F}_{\rm cm}(\bm{r},\bm{r}_c,\bm{Q})=- \zeta_{\rm cm} 
V^{\rm S}\delta\left(\bm{r}-\bm{r}_c\right) \frac{\bm{Q}}{Q}.
\end{align}
At this point we assume that the swarm is perfectly uniform and that therefore
it can be modeled as a periodic lattice. With this assumption then one 
can write $\bm{r}_{c}=c_1 \bm{a}^{(1)}+c_2\bm{a}^{(2)}+c_3\bm{a}^{(3)}$
with, $c_1, c_2, c_3 = 0, \pm 1, \pm 2, ...$. Where $\bm{a}^{(1)}, \bm{a}^{(2)}, \bm{a}^{(3)}$
are the basis vectors determining the unit  cell of the periodic array.
Fig. \ref{fig3} shows a sketch of the periodic array of Janus swimmers.
When writing eq. (\ref{forcecmr}) we have also assumed that the swarm is very large and 
that therefore any disturbances in the periodicity at the edges of it 
can be neglected. \textcolor{black}{Moreover eq. (\ref{forcecmr}) also 
assumes that the swimmers in the swarm do not ``feel'' the concentration 
gradients produced by neighboring swimmers.}
These assumptions allow one to use a method 
proposed by Hasimoto \cite{hasimoto1959periodic}
for the solution of a similar problem, namely  the flow induced by a body force density  
around a periodic array of spheres that is fixed in space.
However, in a swarm the swimmers are not fixed in space, 
they are moving, and therefore
an inertial frame of reference is required in determining the swimmer's
velocity, $\bm{V}_{\rm cm}^{\rm S}$, from the fixed-swimmers solution.
This requires total linear momentum be zero,
which is not true for the fixed swimmers.  However, one can
move the frame of reference from that embedded in the swimmer to a 
swimmer velocity that is relative to the fluid using the momentum conservation equation
\cite{ethier_schieber_2021}. For this problem, the conservation of linear momentum is
$m_c \bm{V}_{\rm cm}^{\rm S}+\rho V_{\rm f} (\bm{V}^{\rm f}+\bm{V}_{\rm cm}^{\rm S})=\bm{0}$.
Where $\bm{V}^{\rm f}$ is the volume-averaged fluid velocity
in the fixed-spheres frame of reference of Hasimoto,
$m_c$ is the mass of a swimmer, $V_{\rm f}$ and 
$\rho$ are the volume and density of the fluid, respectively. 
Note that adding $\bm{V}_{\rm cm}^{\rm S}$
to $\bm{V}^{\rm f}$ in the second term of the linear momentum balance changes
the fluid velocity to an inertial frame of reference.
Then the velocity of the swimmer in the swarm can be calculated as,
$\bm{V}_{\rm cm}^{\rm S}=
-\rho V_{\rm f}/(m_c+\rho V_{\rm f})\bm{V}^{\rm f}=f_{\rm M}\bm{V}^{\rm f}$.
Where $f_{\rm M}$ is the mass fraction of fluid in the swarm.

To calculate the effect of the  the hydrodynamic interactions
with other swimmers on the center of mass 
dynamics of the probe swimmer we start by calculating the 
volume-averaged fluid velocity $\bm{V}^{\rm f}$.
Taking into account the periodicity of the flow field one can use the Fourier series 
to find,
\begin{align}\label{vfieldreal}
\bm{v}(\bm{r})&= \frac{1}{4\pi^2\mu V_0}
\sum_{\bm{k}\neq\bm{0}}\left[\frac{\bm{k} \bm{k}}{k^4}-\frac{\bm{\delta}}{k^2}\right] \cdot 
\bm{F}\left(\bm{k}, \bm{Q}\right)  e^{-2\pi i (\bm{k}\cdot \bm{r})}
\end{align}
where $\bm{k}=c_1 \bm{b}^{(1)}+c_2 \bm{b}^{(2)}+c_3\bm{b}^{(3)}$ are vectors in the 
reciprocal lattice, which satisfy $\bm{k}\cdot\bm{a}^{(j)}=c_j$.
And $V_0=\bm{b}^{(1)} \cdot \left[ \bm{b}^{(2)} \times \bm{b}^{(3)}\right]$ is the volume of the 
unit cell in the physical space. For a simple cubic lattice 
(SCL) $V_0=h^3$, where $h$ is the length 
of the unit cell of the array.
The Fourier transform of the force densities produced by the periodic 
array of swimmers can be written as,
$\bm{F}_{\rm cm}\left(\bm{Q}\right)
=-\zeta_{\rm cm} V^{\rm S}\dfrac{\bm{Q}}{Q}$.
See Section S1 of the
Supplementary Information (SI) for the complete details of the calculation.
% \ref{flowp} 
Then, using the 
volume-average velocity of the fluid, $\bm{V}^{\rm f}$, the velocity of the center 
of mass of a dumbbell in the swarm, $\bm{V}_{\rm cm}^{\rm S}$, 
can be calculated as,
\begin{align}\label{finalvfieldcm}
\bm{V}_{\rm cm}^{\rm S}&=-\frac{f_{\rm M}}{f_{\rm V}V_0}
\int_{V_0} \bm{v}(\bm{r}) d\bm{r}\\ \nonumber
&=-\frac{f_{\rm M} \zeta_{\rm cm} V^{\rm S}}{4\pi^2 \mu f_{\rm V} V_0^2}
\sum_{\bm{k}\neq\bm{0}}\left[\frac{\bm{k} \bm{k}}{k^4}-\frac{\bm{\delta}}{k^2}\right] \cdot 
\frac{\bm{Q}}{Q}\int_{V_0} e^{-2\pi i (\bm{k}\cdot \bm{r})} d \bm{r}.
\end{align}
Where $f_{\rm V}$ is the volume fraction of fluid in the swarm.
 The mathematical details on how one arrives to this expression are given in 
Section S1 of the SI.
%\ref{flowp} 
The sum in eq. (\ref{finalvfieldcm}) is done using the Ewald summation method 
\cite{hasimoto1959periodic, krishnamurthy2015collective} which splits
it into a Fourier space sum and a a real space sum. The Fourier space sum is given 
in eq. (S15) and the real space sum is given in eq. (S16),
%eq. (\ref{finalvfieldcomplexcm})
%eq. (\ref{finalvfieldrealcm})
both in the SI. \textcolor{black}{In Fig. S8 in the SI we 
show a comparison of the results of computing eq. (\ref{finalvfieldcm}) using different 
crystalline lattices for different values of $Q/h$. In particular we have compared the SCL result 
against results obtained with the body-centered cubic lattice (BCL) and with the
face-centered cubic lattice (FCL). The results of 
eq. (\ref{vfieldreal}) using BCL are on average about $7\%$ smaller than the SCL results.
With FCL the difference is larger, and the results are on average about a factor of $1.9$ 
smaller than the SCL result. These differences are in good agreement with the ones
reported by Hasimoto for a similar calculation \cite{hasimoto1959periodic}. 
Here we use the SCL arrangement for all the calculations presented in the main text.}

When comparing the model predictions to a 
real system $f_{\rm M}$ and $f_{\rm V}$ should be calculated using 
the fluid density, $\rho$, the fluid volume, $V^{\rm f}$, 
and the swimmer's mass, $m_c$. However in what follows here we
will assume $f_{\rm M} \approx f_{\rm V}$ and therefore 
$\bm{V}_{\rm cm}^{\rm S}=\bm{V}^{\rm f}$.
In general the result of the sums can be written as, $
\dfrac{\bm{V}_{\rm cm}^{\rm S}}{V^{\rm S}}=
\dfrac{\zeta_{\rm cm}}{2 \pi \mu} 
\dfrac{B_\parallel}{h}\dfrac{\bm{Q}}{Q}$. Where the sub-index 
$\parallel$ indicates parallel to $\bm{Q}$ and the sub-index $\perp$ 
indicates perpendicular to $\bm{Q}$.
The convergence of $B_\parallel$ and $B_\perp$ 
with the number of periodic cells in the crystalline arrangement
is illustrated in Fig. S1 in the SI.
%\ref{figs0}. 
Note that $B_\parallel$ and $B_\perp$ are dimensionless.
The converged values of $B_\parallel$ and $B_\perp$ are given in Figure S7A in the SI.
%\ref{figs6} A.
The sums are performed for a large ensemble of randomly-generated $\bm{Q}$ orientations, 
\textit{i.e.} $\bm{Q}/Q$. The plotted results are averages taken over this 
ensemble and the error bars are the standard errors of the mean. 
Note that $B_\perp=0$ for all values of $Q/h$, which simply means 
that the center of mass always moves in the direction of $\bm{Q}$,
on the other hand $B_\parallel$ can be approximated well by 
\begin{align}\label{Banal}
B_\parallel\approx & 0.175+0.00137\left(\frac{Q}{h}\right)^{-1}-0.888 \left(\frac{Q}{h}\right)
\\ \nonumber
+&2.62 \left( \frac{Q}{h} \right)^2-4.21\left( \frac{Q}{h} \right)^3+
2.69\left( \frac{Q}{h} \right)^4
\end{align}
in the range $0.025 \lesssim (Q/h) \lesssim 0.5$ that is considered here. Note 
that $B_\parallel$ and $B_\perp$ are dimensionless. 

\begin{figure}[h t]
\center
\begin{overpic}[width=\linewidth]{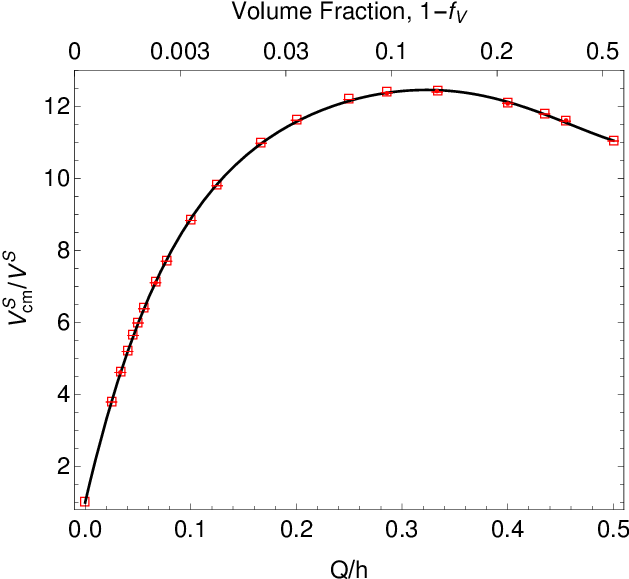}
\end{overpic}
\caption{
Center-of-mass swimming velocity of a microswimmer in 
swarm normalized by the single swimmer velocity,
$V_{\rm cm}^{\rm S}/V^{\rm S}$, as a function of the volume fraction
of swimmers in the swarm. The symbols are the results obtained from 
eq. (\ref{finalvfieldcm}) and the black 
line is the fitted analytic expression given in the text.}
\label{fig1}
\end{figure}

Figure \ref{fig1} shows the velocity of the center 
of mass of the Janus dumbbell in a swarm,  $V_{\rm cm}^{\rm S}$ as 
a function of the volume fraction of swimmers in the swarm, $1-f_{\rm V}$.
To calculate the volume fraction we treat the swimmers as spheres of radius 
$\ell_0/2+R$, where $R$ is the radius of the beads that make up the dumbbell.
We further assume $R=\ell_0/2$ and therefore $1-f_{\rm V}=
\left(\dfrac{4}{3}\pi \ell_0^3\right)/h^3$. 
Note that for low $Q/h$ ratios (or low volume fractions of swimmers) 
the velocity of the center of mass 
of the swimmer increases with increasing volume fraction of swimmers in the swarm. 
Around $Q/h\approx0.32$ or $1-f_{\rm V}\approx0.14$
the velocity $V_{\rm cm}^{\rm S}$ reaches a maximum. This maximum 
velocity achieved by the swimmer in the swarm is about twelve times larger that the velocity 
achieved by a single swimmer in an infinite sea. After the maximum, the velocity
decreases slightly for lower values of $Q/h$. The expression 
for $V_{\rm cm}^{\rm S}$ as a function of the volume fraction of 
microswimmers in the swarm is the first major result presented in this paper.
A subtle but important 
detail of that calculation is the use of the correct (inertial) frame of reference 
for calculating the swimming velocity of the swimmer in the swarm. 

So far we have discussed the effect of hydrodynamic interactions in the
translational velocity of the swimmers in the 
swarm under the assumption of a uniform swarm. In what 
follows we consider the
effect of hydrodynamic interactions on the stability of the swarm. 
In particular we consider the stability around a uniform \textcolor{black}{crystalline} 
state of the swarm. This \textcolor{black}{crystalline arrangement} can be characterized 
by the director vector of the swarm $\bm{n}_0$. In our model $\bm{n}_0$
is a unit vector aligned with one of the basis vectors of the periodic lattice
\textit{i.e.},  $\bm{a}^{(1)}, \bm{a}^{(2)}$ or $\bm{a}^{(3)}$.  
To determine the stability of the swarm we need to calculate perturbations to the
orientation of the swimmers with respect to $\bm{n}_0$.
In order to do this one needs to calculate first the velocity of the
configurational variable, $\bm{Q}$, of the swimmers
in the swarm, $\bm{V}_{\rm conf}^{\rm S}$. 

To calculate $\bm{V}_{\rm conf}^{\rm S}$ 
a higher order approximation of $f(s)$ is required. Here we will use a second-order approximation, which is consistent with our treatment of 
the swimmer as a dumbbell.
Moreover we consider two types of swimming mechanisms.
In the first one the dumbbell is a \textit{pusher} and therefore
all the swimming force acts on
bead $1$, ``the tail'', which pushes bead $2$, 
``the head''. In the second mechanism  the swimming force acts on
bead $2$, ``the head'', which then pulls bead $1$, ``the tail''.
Swimmers that resemble the second mechanism are called 
\textit{pullers}.  The flow field around pushers resembles an 
extensile dipole, since the swimmer pushes fluid back on its 
tail and drags fluid along on its head. On the other hand,
the flow field around the puller resembles a contractile dipole
since the tail drags fluid along and the head pulls fluid back. 
Bacteria such as \textit{Escherichia coli} and Janus microparticles 
with a hemisphere of a platinum catalyst \cite{EbbensLangmuir2011}
have been observed to be pushers. 
Moreover recent experimental work has
investigated the details of the flow field around Pt@SiO$_2$ Janus 
micro colloids \cite{campbell2019experimental}. That work 
has shown that the simplified flow field for a pusher described above 
provides a good approximation for the flow field observed around
the Pt@SiO$_2$ microparticles. In nature 
microorganism with anterior flagella like the
\textit{Chlamydomonas} algae are puller-type swimmers.
The experimental reports of synthetic puller-type
microswimmers are more scarce. Only very recently microparticles 
with a copper catalytic patch,  Cu@SiO$_2$ were shown 
to exhibit puller behavior \cite{sharan2022upstream}.
Here we consider both types of swimmers, pushers and pullers,
to describe the two different swimming mechanisms we use,
\begin{align}\label{fdenQ}
f(s)=\pm \zeta_{\rm conf}V^{\rm S}_{\rm cm}\left[\delta(s-Q/2)-\delta(s+Q/2)\right].
\end{align}
Where $\zeta_{\rm conf}$ is a friction coefficient
and $V_{\rm cm}^{\rm S}=\dfrac{\zeta_{\rm cm}V^{\rm S}}{2 \pi \mu} 
\dfrac{B_\parallel}{h}$ is the velocity of the center of mass calculated above.
Note that the sign in front of $f$
is ``-'' for a pusher and ``+'' for a puller. Note that 
in writing eq. (\ref{fdenQ}) we have treated the dumbbell's beads as points when 
calculating the hydrodynamic forces.
With eq. (\ref{fdenQ}) the force density exerted on the fluid by the 
dumbbell can be written as,
\begin{align}\label{fconf}
\bm{F}_{\rm conf}(\bm{r},\bm{r}_c,\bm{Q})&= \pm\zeta_{\rm conf} V_{\rm cm}^{\rm S}\Bigg\{\delta\left(\bm{r}-\left[\bm{r}_c+\frac{\bm{Q}}{2}\right]\right) \\ \nonumber 
&-\delta\left(\bm{r}-\left[\bm{r}_c-\frac{\bm{Q}}{2}\right]\right)\Bigg\} \frac{\bm{Q}}{Q}.
\end{align}
The sign in front of $\bm{F}_{\rm conf}$ in eq. (\ref{fconf})
is ``+'' for a pusher and ``-'' for a puller. 
As mentioned above, we have accounted for the conservation 
of linear momentum when calculating the effect of hydrodynamic forces 
on the center of mass of the swimmer, in eq. (\ref{finalvfieldcm}). 
Therefore  for calculating 
the effect of hydrodynamic forces on $\bm{Q}$ we use a frame 
of reference embedded in the swimmer and therefore
the net force density produced by eq. (\ref{fconf}) is zero.
The velocity field, $\bm{v}(\bm{r})$ for the force density 
given by  eq. (\ref{fconf}) is again found from the Stokes equation and the 
mass balance for an incompressible solvent, eq. (\ref{stokes_real}). 
The flow field is obtained by using eq. (\ref{vfieldreal})
where in this case the Fourier transform of the force density is,
\begin{align}\label{FConfFou}
\bm{F}_{\rm conf}\left(\bm{k},\bm{Q}\right)&
=\pm\zeta_{\rm conf} V_{\rm cm}^{\rm S}\frac{\bm{Q}}{Q}\left[e^{2\pi i \bm{k} \cdot \bm{Q}/2}-
e^{-2\pi i \bm{k} \cdot \bm{Q}/2}\right].
\end{align}
See Section S1 of the SI for the details of the calculation.
%\ref{flowp}
With eq. (\ref{FConfFou}) the influence of the other swimmers through 
hydrodynamic interaction on the 
configurational degree of freedom of the probe Janus dumbbell is obtained,
\begin{align}\label{finalvfield}
\bm{V}_{\rm conf}^{\rm S}&=\bm{v}\left(\frac{\bm{Q}}{2}\right)-\bm{v}\left(-\frac{\bm{Q}}{2}\right)
\\ \nonumber &=
\pm\frac{\zeta_{\rm conf} V_{\rm cm}^{\rm S}}{\pi ^2 \mu V_0} 
\sum_{\bm{k}\neq\bm{0}}\left[\frac{\bm{k} \bm{k}}{k^4}-\frac{\bm{\delta}}{k^2}\right] \cdot 
\frac{\bm{Q}}{Q} \left(\sin\left[\pi \bm{k}\cdot \bm{Q} \right]\right)^2.
\end{align}
The sign in front of $\bm{V}_{\rm conf}^{\rm S}$ in 
the last expression of eq.~(\ref{finalvfield})
is ``-'' for a pusher and ``+'' for a puller.
Again, the mathematical details on how one arrives to this expression are given in 
Section S1 of the SI.
%\ref{flowp} 
The sum in eq.~(\ref{finalvfield}) is again done using the Ewald summation method 
\cite{hasimoto1959periodic, krishnamurthy2015collective} which splits
it into a Fourier space sum and a a real space sum. The Fourier space sum is given 
in eq.~(S23)  and the real space sum is given in eq.~(S24).
%(\ref{finalvfieldcomplex})
%(\ref{finalvfieldreal})

In eq. (\ref{finalvfield}) we will assume that $\zeta_{\rm conf}=6\pi R \mu$, where $R$ is the radius
of the beads that make up the Janus dumbbell. The velocity 
$\bm{V}_{\rm conf}^{\rm S}$ can 
be decomposed into two distinct parts.
$\left[\bm{V}_{\rm conf}^{\rm S}\right]_\parallel:=
\dfrac{R V_{\rm cm}^{\rm S}A_\parallel}{h} \dfrac{\bm{Q}}{Q}$, which is parallel to $\bm{Q}$,
only causes changes in the length of
the dumbbell. The other part that causes changes in the orientation 
of $\bm{Q}$, and can therefore stabilize or destabilize the swarm can be written as,
$\left[\bm{V}_{\rm conf}^{\rm S}\right]_\perp:=
\dfrac{R V_{\rm cm}^{\rm S}A_\perp}{h} 
\left[\dfrac{\bm{n}_0\times\bm{Q}}{|\bm{n}_0\times\bm{Q}|}\right] \times \dfrac{\bm{Q}}{Q}$. 
Where the sub-index $\perp$ 
indicates perpendicular to $\bm{Q}$ and $\bm{n}_0$ is the initial 
orientation of $\bm{Q}$. 
As pointed out before, $\bm{n}_0$ is chosen to correspond to one 
of the basis vectors of the crystalline arrangement, $\bm{a}^{(1)}$, $\bm{a}^{(2)}$
or $\bm{a}^{(3)}$. Also note that $|\bm{n}_0\times\bm{Q}|$ is the norm 
of the vector $\bm{n}_0\times\bm{Q}$, where ``$\times$'' denotes a cross product.
The function $A_\perp$ is obtained using the relation
\begin{align}\label{Aperp}
A_\perp=\dfrac{h}{R V_{\rm cm}^{\rm S}} 
\left[\dfrac{\bm{Q}}{Q} \times \bm{V}_{\rm conf}^{\rm S}\right]
\cdot\left[\dfrac{\bm{n}_0\times\bm{Q}}{|\bm{n}_0\times\bm{Q}|}\right]
\end{align}
where $\bm{V}_{\rm conf}^{\rm S}$ is calculated numerically using eq. (\ref{finalvfield}).
The results show that the
function $A_\perp$ is positive for pushers but is negative for pullers. 
Note that if $A_\perp$ is negative then $\left[\bm{V}_{\rm conf}^{\rm S}\right]_\perp$ will 
push $\bm{Q}$ in the direction of its initial orientation $\bm{n}_0$, helping to stabilize the swarm. 
But  if $A_\perp$ is positive
it will push $\bm{Q}$ away from $\bm{n}_0$ destabilizing the swarm.

\begin{figure}[h t]
\center
\begin{overpic}[width=\linewidth]{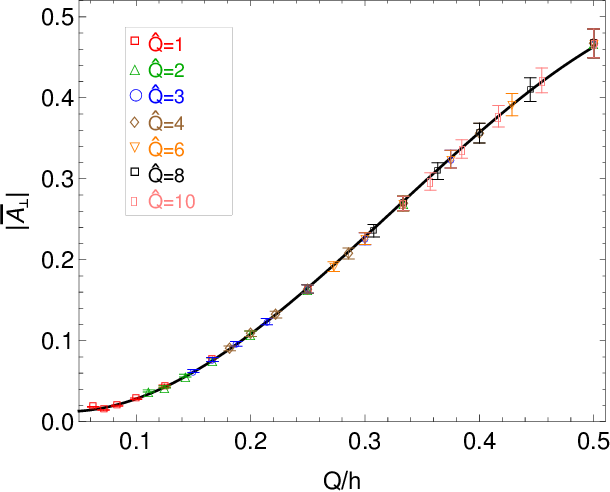}
\put (54,75){\small \textbf{A}}
\end{overpic}\vspace{6mm}
\begin{overpic}[width=\linewidth]{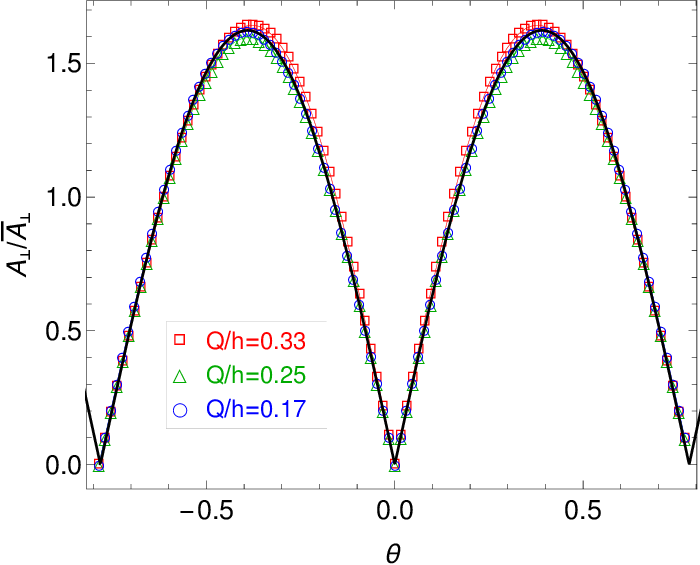}
\put (54,75){\small \textbf{B}}
\end{overpic}
\caption{
(A) Absolute value of the average dimensionless rotational velocity of 
the Janus pusher $| \bar{A}_\perp |$, 
as a function of the ratio between 
its length, $Q$, and the distance from its nearest neighbor in the swarm, $h$.
The average in this case is over the swimmer's orientation.
(B) Orientation dependence of the dimensionless rotational velocity of 
the Janus dumbbell, 
$A_\perp$, defined in eq. (\ref{Aperp}). 
The $x$-axis is the angle between the orientation, $\bm{Q}/Q$,
of the swimmer and its initial orientation, $\bm{n}_0$. Where 
$\bm{n}_0$ is made to coincide with one of the basis vectors of the the crystalline
arrangement. Note that the sign in front of 
 $\bar{A}_\perp$ is positive for pushers but negative for pullers.
 The symbols are the results obtained from eq. (\ref{finalvfield}), the black 
 lines are the fitted analytic expressions given in the text.}
\label{fig1b}
\end{figure}

Also note that $A_\perp$ and $A_\parallel$ are dimensionless.
$A_\perp$ depends both on the ratio $Q/h$ and the angle between 
$\bm{Q}$ and its initial orientation $\bm{n}_0$. For example if $\bm{n}_0=\bm{a}^{(1)}$
and the dumbbell rotates in a plane perpendicular to $\bm{a}^{(3)}$ then the 
angle between $\bm{Q}$ and $\bm{n}_0$ is 
$\theta=\sin^{-1} \left(\left[\bm{a}^{(1)}\times\frac{\bm{Q}}{Q} \right]
\cdot \bm{a}^{(3)}\right)$. The angle $\theta$ is positive when the dumbbell
rotates in a counterclockwise direction and negative when it rotates in a clockwise direction.
$\bar{A}_\perp$ is used to denote
the function $A_\perp$ when its averaged over the orientation dependence.
The convergence of $\bar{A}_\perp$
with the number of cells is illustrated on Figs. S2--S6 in the SI.
%Figs. \ref{figs1}--\ref{figs5}
The converged values of a $\bar{A}_\perp$ are given in Figure \ref{fig1b}A.
As can be seen in Figure \ref{fig1b}A, 
$\bar{A}_\perp$ can be approximated well 
by 
\begin{align}\label{AavAnal}
\bar{A}_\perp\approx\pm\left[0.02-0.4 \left(\frac{Q}{h}\right) + 
5.2 \left(\frac{Q}{h}\right)^2-5.3\left(\frac{Q}{h}\right)^3\right]
\end{align}
in the range $0.04 \lesssim Q/h \lesssim 0.5$ that is considered here.
This analytic function for $\bar{A}_\perp$ is the black
continuous line plotted in Fig. \ref{fig1b}A.
Consistent with what was discussed above,
the sign in front of 
$\bar{A}_\perp$ is positive for pushers and negative for pullers.
The dependence of $A_\perp$ with $\theta$ is shown in 
Figure \ref{fig1b}B, this can be approximated well with the function 
$A_\perp/\bar{A}_\perp\approx|1.6 \sin 4 \theta|$.
This analytic function for $A_\perp/\bar{A}_\perp$ is
the black continuous line plotted in Fig. \ref{fig1b}B.
\textcolor{black}{The results for $A_\parallel$ are given 
in the upper panel of Fig. S7B in the SI. 
As can be seen there, the function $A_\parallel$ 
can be approximated well with the function 
$|A_\parallel|/\hat{h}\approx 6 -3/\hat{Q}$ (see inset in Fig. S7B in the SI). 
Where again $A_\parallel$ is positive for pushers and negative for pullers.
This means that, as expected, pushers feel an extensile hydrodynamic force along $\bm{Q}$,
while pullers feel a contractile hydrodynamic force along $\bm{Q}$. 
This force monotonically increases as the length of the swimmer, $Q$, increases.} 

\textcolor{black}{Note that to obtain eq. (\ref{AavAnal}) we have used the SCL arrangement 
to perform the sums in eq. (\ref{finalvfield}). In Fig. S9 in the SI we 
show comparisons for $Q/h=0.2$ and $Q/h=0.25$ 
of the results of computing eq. (\ref{finalvfield}) using different 
crystalline lattices. In particular we have compared the SCL results 
for $A_\perp$ against results obtained with BCL and FCL. 
Significant differences between the $A_\perp$ calculated using 
these three different lattices are evident
when they are plotted against the angle $\theta$ defined above.  
The differences arise mostly because for SCL the basis vectors are 
orthogonal while for BCL and FCL they are not. Therefore with SCL
the curves of $A_\perp/\bar{A}_\perp~\text{vs.}~\theta$ obtained
by looking at small rotations of $\bm{Q}$ around 
one of the basis vectors of the crystalline lattice are symmetric and universal . 
The universal curve for $A_\perp/\bar{A}_\perp$ obtained with SCL can then be fitted with 
an analytic expression as was done above. In what follows we perform the stability analysis of the 
swarm using the analytic expressions obtained with the SCL arrangement.}

\textcolor{black}{In Figs. \ref{fig1} and \ref{fig1b} we have plotted the predictions of the theory
for volume fractions all the way up to 0.5, and also eqs. (\ref{Banal}) and (\ref{AavAnal})
where fitted to the results in this whole range. But we do not necessarily expect the model
to work well in this whole range. Our exception is that a model
such as the one presented here, which does not account for short-range interactions,
is accurate up to approximately the maximum of the function shown in Fig. \ref{fig1}. 
This maximum occurs at a volume fraction of swimmers of about 0.14
(\textit{i.e.} $Q/h\approx0.32$). Note also that the predictions 
obtained with our simple mean-field model} 
are in good qualitative agreement with recent 
multiparticle simulations that have shown that hydrodynamic 
torques have a stabilizing effect in swarms of pullers while swarms of 
pushers are generally destabilized by hydrodynamic 
interactions \cite{alarcon2017morphology, miles2019active}. 
For instance recently numerical simulations of a
dilute suspension of microswimmers invading a quiescent fluid
showed that sheets of pushers are always unstable while 
the stability of sheets of pullers depends on particle motility \cite{miles2019active}.
In particular that work has shown that for either motile
or immotile pullers, the velocity field can damp the initial
displacement, maintaining the initially aligned order.
However when the motility of the pullers is increased the displacement 
and orientation fields can synchronize, 
leading ultimately to instability and a departure away from the initial ordered state. 
Multiparticle simulations of self-propelled spherical 
particle suspensions in the cluster regime have also shown that
hydrodynamics alone can sustain a cluster phase of active pullers 
\cite{alarcon2017morphology}. In particular the aforementioned work has shown that
in the absence of significant attraction the order decreases algebraically 
with cluster size for pushers, while pullers show a persistent polar order
for all cluster sizes \cite{alarcon2017morphology}.
In the case of strong attraction net polar alignment has been 
observed for clusters of both pushers and pullers \cite{alarcon2017morphology}. 
In other words, that work has shown that
swarms of pusher require strong attraction interactions to be stable while 
swarms of pullers can be stabilized by hydrodynamics alone.
The studies cited above were performed with multiparticle simulations but most of the previous 
mean-field modeling of swarms did not examine how the 
stability depends on the 
volume fraction of swimmer's in the swarm as we do in this work. An exception is the 
work by Subramanian and Koch \cite{subramanian2009critical}. 
But that work used an averaged treatment 
of the hydrodynamic interactions and did not consider the stability around a uniform
\textcolor{black}{crystalline} state as done here.
Therefore, the second important result presented in this paper are the hydrodynamic
torques as function of the volume fraction of swimmers in the swarm shown in Fig. \ref{fig1b}.
So far we have not yet examined the effect of thermal fluctuations
on the stability of the swarm. We did include Brownian forces 
in the stochastic differential 
equations that describe the motion of the swimmer in the swarm, eq. (\ref{SDEs}). 
But so far have not considered their effect on the stability of the swarm,
in the next section we examine their effect.

\section{Brownian forces and stability}\label{results}

As mentioned previously, by a swarm here we mean a uniform solution of swimmers 
that are moving, on average, in the same direction. 
We do not attempt here to describe the emergence of such unifrom 
\textcolor{black}{crystalline arrangement}, we only 
consider its stability. Therefore to simulate a swarm we evolve the dynamics 
of an ensemble of independent swimmers in 
the mean-field described by eqs. (\ref{finalvfieldcm}) and (\ref{finalvfield}).

To study the dynamics of a microswimmer in a swarm we use 
Brownian dynamics (BD) simulations.
To perform the BD simulations we use eq.~(\ref{SDEs})
\textcolor{black}{and the dynamics of fuel consumption are given by 
eqs. (\ref{fuelc}) and (\ref{jrate}). These  equations are first made dimensionless.}
%eq. (\ref{modelfd}).
Lengths are made dimensionless by $\sqrt{\frac{k_{\rm B}T}{k_b}}$ and time 
is made dimensionless by $\lambda_1:=\frac{k_{\rm B}T}{k_bD'_{c\alpha}\gamma'_\alpha}$.
In what follows all dimensionless variables are written with a hat. It
is also useful to write  the net reaction rate in dimensionless form,
$\hat{J}(t)=X_S \left( \frac{X_\rho(t) X_r}{X_r \hat{\ell}_0+1} \right)$.
Where $X_r:=\frac{k'' }{D'_{\alpha\alpha}\gamma_{\alpha\alpha\beta}}
\sqrt{\frac{k_{\rm B}T}{k_b}}$
and $X_S:=\frac{4 \pi R^2}{V}\sqrt{\frac{k_{\rm B}T}{k_b}}$.
Note that $X_r$ is a ratio between two time scales: the time scale
for the mass transfer of reactant from the bulk to the dumbbell,
in the numerator, and in the denominator the
time scale of the reaction at the surface of the catalytic bead. 
Therefore values of $X_r>1$ correspond to an overall reaction rate that is more limited by 
the mass transport of reactants
while values of $X_r<1$ correspond to a reaction rate that is more  
limited by the chemical kinetics.
\textcolor{black}{The dimensionless form 
of all the equations used to perform the BD simulations are
given in Section S2 in the SI, eqs. S25--S28.}

The simulations start from an initial condition in which all the swimmers
in the ensemble are aligned 
with \textcolor{black}{a director vector} $\bm{n}_0$. As previously mentioned, 
$\bm{n}_0$ has to coincide with one of the basis vectors of the periodic lattice.
The analytic expressions for $\bm{V}_{\rm cm}^{\rm S}$ 
and $\bm{V}_{\rm conf}^{\rm S}$ 
that were obtained with the results shown in Figs. \ref{fig1} and \ref{fig1b}
are used in the BD simulations. Moreover, when calculating the hydrodynamic forces
in the BD simulations it is assumed that the length of the dumbbells 
$\hat{Q}$ is always equal to their rest value $\hat{\ell}_0$. Therefore 
the hydrodynamic contractile or extensional force, \textit{i.e.} 
$\left[\bm{V}_{\rm conf}^{\rm S}\right]_\parallel$, is set to zero in the simulations. 
This ``stiffness'' of the swimmers is a common assumption in other kinetic 
theory analyses of the stability of microswimmer suspensions,
moreover for Janus particles and bacteria it is considered a reasonable assumption
\cite{simha2002hydrodynamic,saintillan2008instabilities, 
saintillan2008instabilitiesPF,hohenegger2010stability,subramanian2009critical}.

\textcolor{black}{We have previously shown that 
for the Janus swimmers considered here smaller $\ell_0$ 
produce larger gradients of reactants
along the swimmer's length increasing the swimming velocity \cite{cordoba2020simple}.
Moreover we have also shown that
the time scales at which ballistic motion is observed in the Janus dumbbells 
increases exponentially with $\ell_0$. Here we focus the discussion on the 
effect of the ratio $Q/h$ since this is the more relevant parameter for the 
dynamics and stability of the swarm. 
The hydrodynamic forces that act on the center of mass of the swimmer, eq.  (\ref{Banal}),
and the hydrodynamic torques that act on its orientation, eq. (\ref{AavAnal}), 
depend on the ratio $Q/h$ but not on $Q$ alone.
Therefore for the calculations 
presented in this work we have chosen a value of $\hat{\ell}_0$ that produces 
ballistic motion at short time scales.}

\begin{figure*}[h t]
\center
\includegraphics[width=\linewidth]{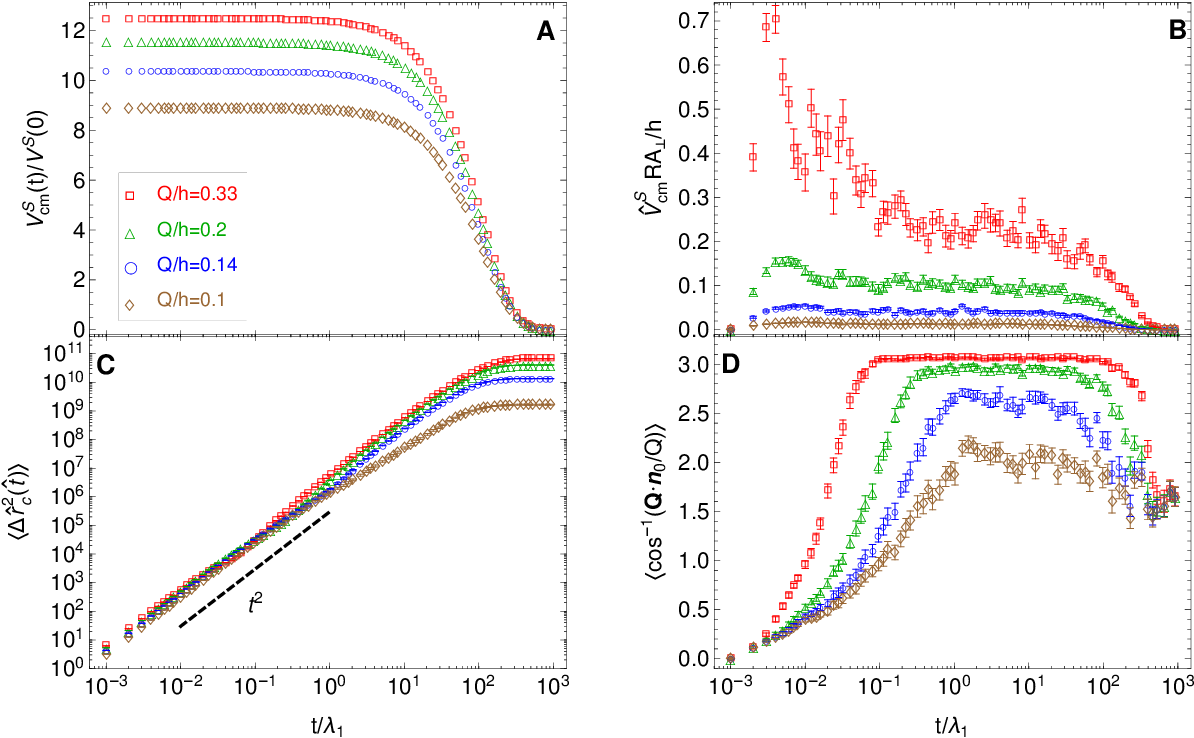}
\caption{Effect of volume fraction of swimmers in the swarm (\textit{i.e.} the ratio $Q/h$)
on the predicted swimming dynamics when the swimmer is treated 
as a \textit{pusher} when calculating hydrodynamic interactions in the swarm.
(A) Hydrodynamic velocity of the center of mass of the Janus dumbbell, $V_{\rm cm}^{\rm S}$. 
(B) \textcolor{black}{Average} hydrodynamic rotational velocity on the end-to-end vector 
of the Janus dumbbell, $R V_{\rm cm}^{\rm S}A_\perp/h$.
(C) Mean-squared displacement of the center of mass of the Janus
swimmer. (D) \textcolor{black}{Average} orientation of the swimmer, $\bm{Q}/Q$, with 
respect to the director vector of the swarm $\bm{n}_0$.
Other parameters are set to, $X_\lambda=1$, $X_r=10$, $X_S=0.01$,
$\hat{\ell}_0=1$, $\hat{R}=0.5$, $\rho_{\alpha}(0)/\overline{\rho}_c=100$.
\textcolor{black}{The error bars in parts B, C and D are standard errors of the mean.}}
\label{fig4}
\end{figure*}

Fig. \ref{fig4} shows the dynamics of a pusher swimmer in a swarm for different 
volume fractions of pusher swimmers in the swarm. In particular, 
Fig. \ref{fig4}A shows the velocity of the center of mass of the swimmer in the 
swarm normalized by the initial single swimmer velocity, $V^{\rm S}_{\rm cm}/V^{\rm S}(0)$.
$V^{\rm S}(0)$ is the velocity of the single swimmer at $t=0$ when the 
fuel concentration is at its maximum. 
In the case of the Janus dumbbell the fuel 
is the reactant $\alpha$, which has an initial average concentration $\rho_\alpha(0)$.
The ratio $V^{\rm S}_{\rm cm}/V^{\rm S}(0)$ 
decreases with time because in the model used here the swimmer dynamics are 
coupled to the fuel dynamics. 
As was pointed out in Section \ref{model} the swimming velocity is 
proportional to the concentration of reactant. 
Therefore as the fuel is consumed the swimming velocity 
of the swimmer decreases. 
The thermodynamically compliant coupling between fuel consumption 
and swimmer motion used here allows us to study more realistic dynamics
instead of an artificially constrained steady-state only.
The ratio $V^{\rm S}_{\rm cm}/V^{\rm S}(0)$  is larger
when the volume fraction of swimmers in the swarm is larger. And the maximum 
value that the ratio $V^{\rm S}_{\rm cm}/V^{\rm S}(0)$ exhibits is 12.4.
The maximum value for this ratio had already been discussed with regards to 
the predictions shown in Fig. \ref{fig1}. 

The rotational velocity of a pusher swimmer 
in the swarm, $R V_{\rm cm}^{\rm S}A_\perp/h$, is shown in Fig. \ref{fig4}B.
First note  this velocity is positive which is what is expected for pushers. 
As discussed previously, positive $A_\perp$ means
that the hydrodynamic torques always rotate the pusher away from its initial orientation $\bm{n}_0$. 
Also note that in general the magnitude of this rotational velocity increases as 
the volume fraction of swimmers in the swarm is increased. The rotational
velocity is initially zero when all the swimmers in the ensemble are perfectly aligned with $\bm{n}_0$
but once thermal forces rotate them slightly away from $\bm{n}_0$
the rotational hydrodynamic velocity grows 
quickly. For all volume fractions of swimmers the rotational velocity reaches a maximum 
at short times, $t/\lambda_1\lesssim10^{-2}$. Then $R V_{\rm cm}^{\rm S}A_\perp/h$
stays at a plateau that remains while the fuel lasts, $10^{-2}\lesssim t/\lambda_1\lesssim10^{2}$. 
After the fuel is depleted, $t/\lambda_1\gtrsim 10^2$, the 
hydrodynamic rotational velocity decays to zero as expected.

To monitor the translational motion of the microswimmers during the BD simulations
the mean-squared displacement of the center of mass,
$\langle \Delta \hat{r}_c^2(t) \rangle=
\langle \left[ \hat{\bm{r}}_c(t)-\hat{\bm{r}}_c(0) \right]^2 \rangle$ 
is also calculated, where the brackets, $\langle ...  \rangle$, denote an ensemble average. 
$\langle \Delta \hat{r}_c^2(t) \rangle$ for a pusher swimmer 
in a swarm is shown in Fig. \ref{fig4}C.
For the parameter values used in this paper the motion of the center of mass is 
clearly ballistic from the beginning and remains in that 
regime until the fuel is depleted. However,
as has been previously shown $\langle \Delta \hat{r}_c^2(t) \rangle$ can exhibit up 
to four characteristic regions \cite{cordoba2020simple}.
A diffusive region, $\langle \Delta \hat{r}_c^2(t) \rangle \sim t^1$, can observed at short times. 
This can be followed by a ballistic region where $\langle \Delta \hat{r}_c^2(t) \rangle\sim t^{2}$.
The ballistic region can be followed by 
another diffusive region where the diffusivity of 
the dumbbell is larger than in the first diffusive region by a factor 
that can range from 5 to 17. This second diffusive region 
appears at times at which the concentration of reactant and 
the swimming velocity are still somewhat large. However due to tumbling,
{\it i.e.} change in the direction of $\bm{Q}$ due to Brownian forces, the
swimmer can not maintain the ballistic motion for a longer period of time.
Following this second diffusive region a plateau can be observed in 
$\langle \Delta \hat{r}_c^2(t)$.
This region arises when the concentration of reactant is very small, 
the reactant is near depletion but not quite yet depleted. 
For very small but still finite concentration of reactant 
the center of mass, $\bm{r}_c(t)$, is still coupled to the orientation of 
the dumbbell, $\bm{Q}/Q$. At this point while $\bm{Q}/Q$
keeps changing due to Brownian forces the center of mass, $\bm{r}_c(t)$, moves very little 
in every new direction of $\bm{Q}$, so effectively the dumbbell appears trapped.
When the reactant is completely consumed, \textit{i.e.} $\rho_{\alpha}=0$, 
the dynamics of the center of mass, $\bm{r}_c(t)$, become completely decoupled from 
the $\bm{Q}(t)$ dynamics and a final diffusive region of 
purely thermal nature is observed. A more extensive analysis of
the translational and rotational dynamics of the Janus dumbbell model used here can be found
in our previous work \cite{cordoba2020simple}. In this paper, to focus on the 
interpretation of the effects of hydrodynamic interactions
we have chosen parameters that maximize the duration
of the ballistic motion of the dumbbell. 
In Fig. \ref{fig4}C it can be observed that in a swarm the magnitude 
of $\langle \Delta \hat{r}_c^2(t) \rangle$ increases with
increasing volume fraction of swimmers in the swarm. This is
a direct consequence of the dependence of $V^{\rm S}_{\rm cm}$
on the volume fraction of swimmers in the swarm.

\begin{figure*}[h t]
\center
\includegraphics[width=\linewidth]{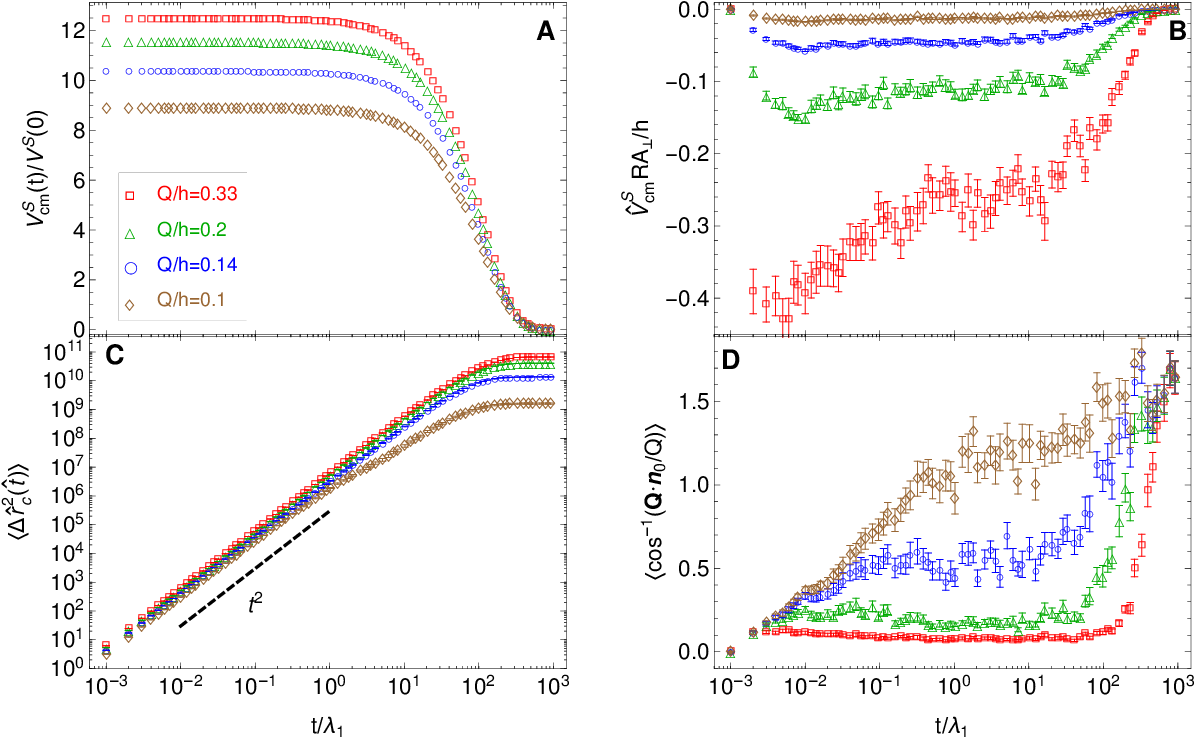}
\caption{Effect of volume fraction of swimmers in the swarm (\textit{i.e.} the ratio $Q/h$)
on the predicted swimming dynamics when the swimmer is treated 
as a \textit{puller} when calculating hydrodynamic interactions in the swarm.
(A) Hydrodynamic velocity of the center of mass of the Janus dumbbell, $V_{\rm cm}^{\rm S}$. 
(B) Hydrodynamic rotational velocity on the end-to-end vector 
of the Janus dumbbell, $R V_{\rm cm}^{\rm S}A_\perp/h$.
(C) Mean-squared displacement of the center of mass of the Janus
swimmer.  (D) Orientation of the swimmer, $\bm{Q}/Q$, with 
respect to the director vector of the swarm $\bm{n}_0$.
Other parameters are set to, $X_\lambda=1$, $X_r=10$, $X_S=0.01$,
$\hat{\ell}_0=1$, $\hat{R}=0.5$, $\rho_{\alpha}(0)/\overline{\rho}_c=100$.
\textcolor{black}{The error bars in parts B, C and D are standard errors of the mean.}}
\label{fig5}
\end{figure*}

Fig. \ref{fig4}D shows the \textcolor{black}{average} angle between the orientation 
of the swimmer and the director vector of the swarm, 
$\langle \cos^{-1}\left(\bm{Q}\cdot\bm{n}_0/Q\right)\rangle$, 
as a function of time. This figure is for a swarm of pusher swimmers. 
\textcolor{black}{Note that the angle between $\bm{Q}$ and $\bm{n}_0$ can vary 
between $0$ and $\pi$ radians.} This angle starts at zero at $t=0$ because initially all the 
swimmers in the ensemble are aligned with 
the director vector $\bm{n}_0$. In this case of pushers the angle grows quickly once 
Brownian forces start to rotate the swimmers slightly away from their initial orientation. 
In agreement with what is observed for $R V_{\rm cm}^{\rm S}A_\perp/h$ the 
angle grows faster for larger volume fractions of swimmers in the swarm. 
Also similarly to what is observed for the rotational velocity the angle reaches a plateau
that remains while the fuel last. This plateau occurs at larger angles for larger volume 
fractions of swimmers in the swarm. After the fuel is depleted the angle between 
$\bm{Q}$ and $\bm{n}_0$ starts to decrease again. 
\textcolor{black}{When the angle is unbiased by hydrodynamic torques
(\textit{i.e.} rotations occurring due to Brownian forces only), 
which is what happens when the fuel is depleted, the distribution becomes
uniform. And therefore the value of 
$\langle \cos^{-1}\left(\bm{Q}\cdot\bm{n}_0/Q\right)\rangle$ 
converges to the mean between $0$ and $\pi$ radians, which is $\pi/2$ radians.}
In summary this plot 
of the relative orientation of the swimmers clearly shows that the swarm of 
pusher microswimmers  is always unstable. Moreover it becomes unstable faster
when the volume fraction of swimmers in the swarm is larger.

In Fig. \ref{fig5} the dynamics of a swarm of pullers are shown as a function 
of the volume fraction of swimmers in the swarm. Again Fig. \ref{fig5}A
shows the velocity of the center of mass of the swimmer in the 
swarm normalized by the initial single swimmer velocity, $V^{\rm S}_{\rm cm}/V^{\rm S}(0)$.
This plot is identical to Fig. \ref{fig4}A because $V^{\rm S}_{\rm cm}$ is the same 
for pullers and pushers. The mean-squared displacement of the center of mass of the 
puller swimmers in the swarm is shown in Fig. \ref{fig5}C. 
As for pushers this \textcolor{black}{$\langle \Delta \hat{r}_c^2(t)\rangle$}
displays ballistic motion while the fuel concentration remains high, in general
the behavior of $\langle \Delta \hat{r}_c^2(t) \rangle$ for 
pullers is identical to the one already shown for pushers in Fig. \ref{fig4}B.

Fig. \ref{fig5}B shows the rotational velocity of a puller swimmer 
in the swarm, $R V_{\rm cm}^{\rm S}A_\perp/h$.
Differently from the velocities in Fig. \ref{fig4}B this predictions for pullers 
show that the rotational velocities for pullers are negative.
Which means that the hydrodynamic torques always rotate the puller 
towards its initial orientation $\bm{n}_0$. 
Also note that in general the magnitude of this negative rotational velocity increases as 
the volume fraction of swimmers in the swarm is increased. The rotational
velocity is initially zero when all the swimmers in the ensemble are perfectly aligned with $\bm{n}_0$
but once thermal forces start to rotate them slightly away from $\bm{n}_0$ 
the magnitude of the negative rotational hydrodynamic velocity grows 
quickly. For all volume fractions of swimmers the
magnitude of the negative rotational velocity reaches a maximum 
at short times, $t/\lambda_1\lesssim10^{-2}$. Then $R V_{\rm cm}^{\rm S}A_\perp/h$
remains at a plateau that remains while the fuel lasts, $10^{-2}\lesssim t/\lambda_1\lesssim10^{2}$. 
After the fuel is depleted, $t/\lambda_1\gtrsim 10^2$, the 
hydrodynamic rotational velocity returns to zero as expected.

\begin{figure}[h t]
\center
\includegraphics[width=\linewidth]{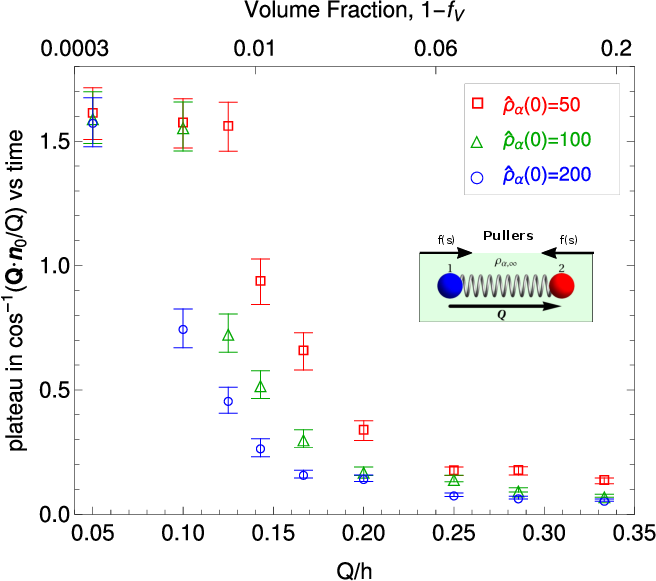}
\caption{Plateau in the angle between $\bm{Q}$ and $\bm{n}_0$
in a swarm of puller microswimmers
as a function of the volume fraction of swimmers in the swarm.
Results for different initial concentration of fuel are shown. Larger 
$V^{\rm S}(0)$ corresponds to a larger initial concentration of fuel.
Large values of the plateau plotted here indicate unstable swarms.}
\label{fig6}
\end{figure}

Fig. \ref{fig5}D shows the \textcolor{black}{average angle} between the orientation of the swimmers
and the director vector of the swarm, 
$\langle \cos^{-1}\left(\bm{Q}\cdot\bm{n}_0/Q\right)\rangle$, 
as a function of time. But in this figure the predictions are for a swarm of puller swimmers. 
Again, this angle starts at zero at $t=0$ because initially all the swimmers in the 
ensemble are aligned with 
the director vector $\bm{n}_0$. But differently from the case of pushers shown in 
Fig. \ref{fig4}D, in this case of pullers the angle grows slightly due to
thermal fluctuations but the growth is damped by the hydrodynamic torques. 
In agreement with what is observed for $R V_{\rm cm}^{\rm S}A_\perp/h$ the 
angle grows faster for smaller volume fractions of swimmers in the swarm. That is 
because for those lower volume fractions the negative hydrodynamic torques 
are smaller in magnitude. \textcolor{black}{
The angle $\langle \cos^{-1}\left(\bm{Q}\cdot\bm{n}_0/Q\right) \rangle$}
reaches a plateau
that remains while the fuel last. This plateau occurs at smaller angles for larger volume 
fractions of swimmers in the swarm, when the 
magnitude of the stabilizing hydrodynamic torques are larger.
After the fuel is depleted the angle between 
$\bm{Q}$ and $\bm{n}_0$ starts to increase at a fast rate due to Brownian forces
and the fact that the
magnitude of the  hydrodynamic torques is not longer large enough to stabilize the swarm. 

The value that the \textcolor{black}{angle 
$\langle \cos^{-1}\left(\bm{Q}\cdot\bm{n}_0/Q\right)\rangle$}
reaches at the plateau 
observed in plots such as the one shown in Fig. \ref{fig5}D are shown as a function 
of volume fraction of swimmers in the swarm in Fig. \ref{fig6}. Note that the 
angle always stays below 0.2 radians for volume fractions larger than 0.06. 
Moreover Fig. \ref{fig6}
also shows how the value of this angle depends on $\rho_{\alpha}(0)$. Which is 
the the initial average concentration of fuel in the system.  
For the larger initial concentrations
of fuel the angle between $\bm{Q}$ and $\bm{n}_0$ stays below 0.2 radians 
for volume fractions as low as 0.02. This is because at higher initial fuel concentrations 
the hydrodynamic torques reach larger values and therefore are able to stabilize the 
uniform \textcolor{black}{crystalline} 
swarm of pullers at lower volume fractions of swimmers. For volume fractions 
lower than 0.02 the swarm of pullers is unstable due to thermal fluctuations. 
The possibility of generating phase diagrams such as the one shown in Fig. \ref{fig6}
is the third major result of this paper. With the model 
presented here these types of phase diagrams are relatively cheap to generate and can be useful 
in determining the experimental conditions at which stable swarms of microswimmers 
can be realized and observed.

\section{Conclusions}\label{conclusions}

In this paper a thermodynamically compliant microswimmer model
was proposed to describe swarms of microswimmers. 
The model is based on 
our previously introduced single-swimmer model \cite{cordoba2020simple}.
In particular,
the swimming velocity of a swimmer in a uniform swarm as function of the 
volume of swimmers in the swarm was calculated. 
A microswimmer in the swarm can 
swim up to twelve times faster than when swimming alone in an infinite sea,
when the volume fraction of swimmers in the swarm is about 0.15.
Moreover we have also studied the stability of a swarm 
around a \textcolor{black}{uniform crystalline} state.
We have done this by calculating the hydrodynamic torques generated by the 
swimmers as a function of the volume fraction of swimmers in the swarm.
The predictions presented here are in good agreement with recent 
multiparticle simulations that have shown that hydrodynamic 
torques have a stabilizing effect in swarms of pullers while swarms of 
pushers are generally destabilized by hydrodynamic 
interactions \cite{alarcon2017morphology, miles2019active}.
Moreover differently from most of the previous mean-field modeling of microswimmer
swarms we have accounted for high order hydrodynamic interactions between 
microswimmers. This allowed us to study the swarm's stability as function of the 
volume fraction of swimmers in the swarm. 

Although the model proposed here accounts for high-order hydrodynamic interactions 
it is still a dilute theory in that it does not capture close or steric interactions 
that may be important in determining the emergence
of collective swimming states. Near-field hydrodynamic 
interactions can lead to complex dynamics of attraction or repulsion
but do not change the far-field flow field. \textcolor{black}{
We expect that for volume fractions of swimmers below 0.14 (\textit{i.e.}
$Q/h\approx0.32$) the treatment of the hydrodynamics used here should be accurate.} 
In the case of higher volume fractions some theoretical progress has been 
achieved by studying the near body interactions of two swimmers
\cite{pooley2007hydrodynamic, ishikawa2006hydrodynamic, ziegler2021theoretical,
guan2023squirmers}. Our model also assumes that the swimmers in the 
swarm do not ``feel'' the concentration gradients caused by neighboring swimmers. 
This assumption is expected to be valid for dilute swarms, but it 
can also break down at large volume fractions of swimmers.
For example simulations of dense clusters of microswimmers
have shown that an individual swimmer
will respond to both its self-generated concentration gradient
and that due to other swimmers
\cite{colberg2017many,huang2017chemotactic,huang2016microscopic}.

We have also examined the stability of swarms of pushers and pullers 
\textcolor{black}{starting from a crystalline arrangement and}
in the presence of thermal fluctuations. We find that hydrodynamic torques 
can stabilize swarms of pullers. We observe stable swarms of
pullers at volume fractions of swimmers in the swarm above 0.02. 
In the case of pushers the hydrodynamic torques accelerate the destabilization
of the swarm that is initialized by the thermal fluctuations.
These predictions are in good qualitative agreement with recent 
multiparticle simulations that have shown that hydrodynamic 
torques have a stabilizing effect in swarms of pullers while swarms of 
pushers are generally destabilized by hydrodynamic 
interactions \cite{alarcon2017morphology, miles2019active}.
Experimental observation on the ``collective behavior'' of 
moderately concentrated suspensions microswimmers is only available 
for pusher swimmers such as \textit{Escherichia coli}
\cite{wu2000particle} and \textit{Bacillus subtilis} \cite{sokolov2007concentration, dunkel2013fluid}.
Since swarms of pusher swimmers are unstable that work has focused mostly 
on the enhance mixing that is achieved by the complex flow patterns formed 
by the unstable swarms. The model presented here may facilitate the design 
of experiments with puller swimmers in which \textcolor{black}{highly ordered} 
swarms could be observed and studied.

To conclude, is important to emphasize that the theory presented here
is a mean-field model, where most of the calculations are done analytically. 
Simulations are preformed only 
to study the effect of Brownian forces on the stability of the swarm. But even when simulations 
are necessary, it is an ensemble of independent probe swimmers in a 
mean-field that are simulated so parallelization is trivial and ideal. 
Predictions with a mean-field theory like the one presented here
are cheaper and faster to obtain than with multiparticle simulations.
Most of the previous mean-field models of swarms did 
not include hydrodynamic interactions between swimmers while our theory 
does account for far-field hydrodynamic interactions accurately.
Although the crystalline arrangement of the swarm assumed here is clearly 
idealized, it still gives useful insight into the role of hydrodynamic interactions in a swarm. 
Therefore our model can provide valuable guidance 
for the design or interpretation of experiments
before resorting to more detailed and expensive simulations. 
Also, the theory presented here can be used to test multiparticle simulations in certain 
limits where agreement with our mean-field theory should be expected.

\section*{Supplementary Material}
\noindent Additional mathematical details and plots 
used to obtain the equations for the model used here
are given in the supplementary information file.

\section*{Data Availability Statement}
The data that supports the findings of this study are available 
within the article [and its supplementary material].

%\section*{Author contributions statement}
%\noindent A.C. and J.D.S. conceived the project. A.C. conducted the research and wrote the first 
%draft of the paper. J.D.S. helped with interpretation and analysis of results. 
%A.C. and J.D.S. wrote and reviewed the paper.
%
%\section*{Competing interests}
%\noindent The authors declare no competing interests.
%
%\section*{Data availability}
%\noindent All data generated or analysed during this study are included in 
%this published article and its supplementary information files.

\section*{References}

%\nocite{*}

\bibliography{referencesswimmer}% Produces the bibliography via BibTeX.

\end{document}